\documentclass[10pt,twocolumn,letterpaper]{article}
\usepackage[utf8]{inputenc}
\usepackage[margin=0.8in]{geometry}
\usepackage{setspace}
\usepackage{graphicx}
\usepackage{booktabs}
\usepackage{amsmath}%
\usepackage[labelfont=bf,skip=5pt,justification=justified, format=plain]{caption}
\usepackage[caption=false]{subfig}
\usepackage[style=chem-acs,articletitle=true,maxbibnames=30]{biblatex}
\usepackage [english]{babel}
\usepackage [autostyle, english = american]{csquotes}
\MakeOuterQuote{"}
\usepackage{lipsum}
\usepackage{url}

\begin{document}

\begin{singlespace}

\title{High Temperature Ferromagnetism in Cr$_{1+x}$Pt$_{5-x}$P}

\author{Tyler J. Slade,$^{1,2*}$, Nao Furukawa,$^{1,2}$ Tanner R. Smith,$^{1,2}$ Juan Schmidt,$^{1,2}$ \\  Ranuri S. Dissanayaka Mudiyanselage,$^3$ Lin-Lin Wang,$^{1}$ Weiwei Xie,$^{3,4}$ Sergey L. Bud’ko,$^{1,2}$ \\ Paul C. Canfield$^{1,2*}$}
\date{}

\twocolumn[
\begin{@twocolumnfalse}

\maketitle

\begin{center} 
    
\textit{$^{1}$Ames National Laboratory, US DOE, Iowa State University, Ames, Iowa 50011, USA} \\  
\textit{$^{2}$Department of Physics and Astronomy, Iowa State University, Ames, Iowa 50011, USA} \\
\textit{$^{3}$Department of Chemistry and Chemical Biology, The State University of New Jersey Rutgers, Piscataway, NJ 08854, USA} \\
\textit{$^{4}$Department of Chemistry, Michigan State University, East Lansing, MI, 48824, USA} \\

\begin{abstract}

We present the growth and basic magnetic and transport properties of Cr$_{1+x}$Pt$_{5-x}$P. We show that single crystals can readily be grown from a high-temperature solution created by adding dilute quantities of Cr to Pt-P based melts. Like other 1-5-1 compounds, Cr$_{1+x}$Pt$_{5-x}$P adopts a tetragonal \textit{P}4/\textit{mmm} structure composed face-sharing CrPt$_3$ like slabs that are broken up along the \textit{c}-axis by sheets of P atoms. EDS and \textit{X}-ray diffraction measurements both suggest Cr$_{1+x}$Pt$_{5-x}$P has mixed occupancy between Cr and Pt atoms, similar to what is found in the closely related compound CrPt$_3$, giving real compositions of Cr$_{1.5}$Pt$_{4.5}$P (\textit{x} = 0.5). We report that Cr$_{1.5}$Pt$_{4.5}$P orders ferromagnetically at \textit{T}$_C$ = 464.5 K with a saturated moment of $\approx$ 2.1 $\mu_{\textit{B}}$/Cr at 1.8 K. Likely owing to the strong spin-orbit coupling associated with the large quantity of high \textit{Z}, Pt atoms, Cr$_{1.5}$Pt$_{4.5}$P has exceptionally strong planar anisotropy with estimated anisotropy fields of 345 kOe and 220 kOe at 1.8 K and 300 K respectively. The resistance of Cr$_{1.5}$Pt$_{4.5}$P has a metallic temperature dependence with relatively weak magnetoresistance. Electronic band structure calculations show that CrPt$_5$P has a large peak in the density of states near the Fermi level which is split into spin majority and minority bands in the ferromagnetic state. Furthermore, the calculations suggest substantial hybridization between Cr-3d and Pt-5d states near the Fermi level, in agreement with the experimentally measured anisotropy.

\end{abstract}

\end{center}

\vspace{7mm}

\end{@twocolumnfalse}
]

\section{Introduction}

From both an applied and fundamental perspective, ferromagnetism continues to provide opportunities and challenges for new research. At a practical level, the high cost and environmental concerns associated with the mining and purification of rare earth elements necessitates the search for new rare earth free permanent magnets for applications ranging from energy production, memory storage, and high performance electric engines to emerging quantum technology such as spintronics.\autocite{RevModPhys.63.819,6008648,weller2000high,victora2005exchange,shiroishi2009future,felser2007spintronics} More fundamentally, it remains challenging to accurately predict the transition temperatures and even the type of magnetic order observed in many materials, especially in metallic compounds where the magnetism originates from the itinerant electrons rather than local moments.\autocite{stoner1938collective,wohlfarth1951xliii,landrum2000orbital,takahashi2013spin} In such itinerant ferromagnets, the physics and possible emergent phases associated with suppressing the Curie temperature towards 0 K using pressure, and/or chemical doping is less understood than for antiferromagnets and remains active frontier in condensed matter research.\autocite{RevModPhys.88.025006,kaluarachchi2017tricritical,PhysRevB.103.075111,PhysRevLett.124.147201,PhysRevB.103.054419,shen2020strange} These factors, amongst others, make the discovery of new, transition metal based, ferromagnetic metals highly desirable.

Recently, several new materials in the Mn(Pt,Pd)$_5$Pn (Pn = P, As) family, referred to here as 1-5-1 compounds, were discovered and found to order ferromagnetically near room temperature. Specifically, MnPt$_5$As exhibits ferromagnetic order below \textit{T}$_{\text{C}}$ $\approx$ 280 K and MnPd$_5$P below $\approx$ 295 K.\autocite{gui2020crystal,Slade_Canfield_ZAAC} While not ferromagnetic, the isostructural materials MnPt$_5$P and FePt$_5$P also have interesting magnetic properties, as MnPt$_5$P orders antiferromagnetically at 192 K (with a small ferromagnetic, \textit{q} = 0 component),\autocite{gui2020chemical,Slade_Canfield_ZAAC} and FePt$_5$P is an itinerant antiferromagnet with three closely spaced transitions between 70 K and 90 K.\autocite{gui2021spin,Slade_Canfield_ZAAC}

The magnetic 1-5-1 compounds are part of the much larger \textit{X}(Pt,Pd)$_5$Pn family that crystallizes in the tetragonal \textit{P}4/\textit{mmm} ($\#$123) space group and where \textit{X} can be a late transition metal (Ag, Zn, Cd, Hg), a B-group element (Al--Tl), or a 3d transition metal (Mn, Fe).\autocite{el1970verzerrte,zakharova2018ternary} The \textit{X}(Pt,Pd)$_5$Pn materials are isostructural to the well-known family of heavy fermion superconductors Ce\textit{T}In$_5$ (\textit{T} = Co, Rh, Ir),\autocite{PhysRevLett.84.4986,petrovic2001heavy,petrovic2001new,movshovich2001unconventional} and similarly to how Ce\textit{T}In$_5$ can be pictured as being derived from CeIn$_3$, a useful way to describe the \textit{X}(Pt,Pd)$_5$Pn crystal structure is as a quasi-two dimensional material formed by breaking up the parent cubic compound \textit{X}Pt$_3$ into \textit{X}-Pt layers that are separated by sheets of P atoms along the \textit{c}-axis (see Figure \ref{Structure_PXRD}a).

From the standpoint of materials discovery, the above picture is attractive given that nearly all \textit{X} elements that are reported to form ternary \textit{X}(Pt,Pd)$_5$Pn compounds are also reported to form \textit{X}Pt$_3$ binaries. With this in mind, identifying known \textit{X}Pt$_3$ for which a corresponding \textit{X}Pt$_5$Pn is not yet reported may be a fruitful guide to discovering new magnetic materials. Here we focus on \textit{X} = Cr. CrPt$_3$ is a \textit{T}$_{\textit{C}}$ $\approx$ 490 K ferromagnetic with among the highest magnetic anisotropy energies known for a cubic material.\autocite{pickart1963neutron,besnus1973magnetic,maruyama1995separation,leonhardt1999crpt,goto1977effects} A high anomalous hall effect of 1750 S$\cdot$cm$^{-1}$ was recently measured on CrPt$_3$ films and attributed to the Berry curvature stemming from two gaped nodal lines near the Fermi level.\autocite{markou2021hard} Based on the unique properties exhibited by CrPt$_3$, we reasoned that a hypothetical CrPt$_5$P may also host interesting behavior.

Here, we report the discovery and basic physical properties of the compound Cr$_{1+x}$Pt$_{5-x}$P. We find that single crystals can readily be grown from solution by adding Cr into melts based on the single phase liquid region above the Pt--P eutectic composition.\autocite{Slade_Canfield_ZAAC} Cr$_{1+x}$Pt$_{5-x}$P is the first ternary compound in the Cr--Pt--P phase space and adopts the same tetragonal \textit{P}4/\textit{mmm} crystal structure as the other known 1--5--1 materials. Our EDS and X-ray diffraction characterization suggests mixed occupancy between Cr and Pt atoms, giving true compositions of Cr$_{1+x}$Pt$_{5-x}$P with \textit{x} between 0--0.5, which is similar to what is observed in the CrPt$_3$ parent compound. The samples characterized here have \textit{x} $\approx$ 0.5, and we show that Cr$_{1.5}$Pt$_{4.5}$P orders ferromagnetically below \textit{T}$_{\text{C}}$ = 464.5 K with a saturated moment of 2.1 $\mu_{\textit{B}}$/Cr at 1.8 K. The \textit{ab}-plane is the easy direction and the magnetic anisotropy of Cr$_{1.5}$Pt$_{4.5}$P is exceptionally high, with an estimated anisotropy fields of $\approx$ 345 kOe and 220 kOe at 1.8 K and 300 K respectively. Transport measurements indicate typical metallic behavior with a moderate negative magnetoresistance when the field is applied within the \textit{ab}-plane. Density functional theory calculations show a strong peak in the density of states near the Fermi level which is split into majority and minority spin bands in the magnetic state, consistent with the picture for itinerant magnetism. Likewise, the calculations show significant hybridization between Cr-3d and Pt-5d states, which likely underpins the strong magnetic anisotropy.

\section{Experimental Details}

\subsection{Crystal Growth}
The staring materials were elemental Pt powder (Engelhard, 99+$\%$ purity), red P (Alpha Aesar, 99.99$\%$) and Cr pieces (Alpha Aesar, 99.999$\%$). The elements were weighed according to a nominal molar ratio of Cr$_{9}$Pt$_{71}$P$_{20}$ and contained in the bottom side of a 3-piece alumina Canfield crucible set (CCS, sold by LSP Ceramics )\autocite{canfield2016use,lspceramics_CCS} The CCS was sealed in a fused silica ampule and was held in place with a small amount of silica wool, which serves as cushioning during the decanting step.\autocite{canfield2019new} The ampules were evacuated three times and back-filled with $\approx$ 1/6 atm Ar gas prior to sealing. 

The crystal growth took place over three steps. First, using a box furnace, the ampule was slowly warmed to 250°C over 6 h, then to 1180°C in an additional 8-12 h. After dwelling at 1180°C for 6 h, the furnace was cooled to 950°C over 40 h, after which the remaining liquid phase (a Cr-depleted, Pt-P-rich melt) was decanted by inverting the ampule into a centrifuge with specially made metal rotor and cups.\autocite{canfield1992growth,canfield2016use,canfield2019new} After cooling to room temperature, the ampule was opened and the products found to be metallic, highly inter-grown, block-like crystals. 

The remaining solidified liquid phase that was decanted and captured in the "catch" crucible of the CCS was next reused in a second step.\autocite{Slade_Canfield_ZAAC} Here, a new 3-piece CCS was assembled, using the decanted liquid in the new "growth" end. The crucible set was sealed in an ampule as described above and heated rapidly (in $\approx$ 2 h) to 975°C. After holding for 6 h, the furnace was slowly cooled to 800°C over 75 h and the remaining liquid again decanted. The ampule was opened to reveal a mixture of block-like crystals like those described above as well as thin, nicely formed metallic plates. In the final step, the remaining solidified liquid phase in the "catch" crucible was again reused. After evacuating and sealing into an ampule, the furnace was heated to 825°C, held for 6 h, and then cooled to 700°C over 60 h where the excess liquid was decanted. The final step yielded only thin, plate-like crystals like those also found in the second step. Pictures of several plates are shown in the inset to Figure \ref{Structure_PXRD}c. 

Analysis with EDS suggested that both the block-like crystals and thinner plates belonged to the same phase, with a putative chemical formula of Cr$_{1+x}$Pt$_{5-x}$P (see discussion section for details). This indicates that the initial, more Cr-rich melts decanted at higher temperatures produce thicker block-like crystals, whereas using relatively Cr-depleted melts and decanting at lower temperatures results in the the thin plates. Both the three-dimensional blocks and plates were malleable, and the plates could be bent and deformed using tweezers. Furthermore, the plates could be cleaved within the basal plane. Because the block-like crystals were generally inter-grown with less clear orientation, the well formed, plate crystals were used in all following measurements.

\subsection{Crystal Structure Determination}

We initially attempted to use single crystal X-ray diffraction to determine the crystal structure of the plate-like crystals. Samples were analyzed with a Bruker D8 Quest Eco single-crystal X-ray diffractometer equipped with a Photon II detector and Mo radiation ($\lambda$K$_{\text{$\alpha$}}$ = 0.71073 \AA). As noted above, the crystals are malleable and readily deformed, and as a result, selecting a suitable piece was challenging and the reflections were always weak and tailed (this is likely compounded by the large quantity of Pt in the system, which is strongly absorbing). While the structure refinements were consistent with the \textit{P}4/\textit{mmm} 1-5-1 arrangement\autocite{gui2020chemical,gui2020crystal,gui2021spin} the poor data quality rendered assessment of the structural parameters (site occupancy, thermal parameters, etc.) impossible.

To better determine the phase and assess the purity of the crystals, samples were analyzed with powder X-ray diffraction (PXRD). A representative selection of $\approx$5-10 plate-like crystals was ground to a fine powder, sifted through a 33 micron mesh sieve, and analyzed using a Rigaku Miniflex-II instrument operating with Cu-K$\alpha$ radiation with $\lambda$ = 1.5406 Å (K$\alpha_1$) and 1.5443 (K$\alpha_2$) Å at 30 kV and 15 mA. The powder patterns were refined using GSAS-II software.\autocite{toby2013gsas} 

\subsection{Chemical Composition}
The chemical composition was determined by energy dispersive X-ray spectroscopy (EDS) quantitative chemical analysis using an EDS detector (Thermo NORAN Microanalysis System, model C10001) attached to a JEOL scanning-electron microscope (SEM). The measurements were performed at three different positions on each crystal's face (perpendicular to \textit{c}-axis), revealing good homogeneity in each crystal. An acceleration voltage of 16 kV, working distance of 10 mm and take off angle of 35$^{\circ}$ were used for measuring all standards and crystals with unknown composition. A MnPt$_5$P single crystal was used as a standard for Pt and P quantification, and a LaCrGe$_3$ single crystal was used as a standard for Cr. The spectra were fitted using NIST-DTSA II Microscopium software.\autocite{newbury2014rigorous} The average compositions and error bars were obtained from these data, accounting for both inhomogeneity and goodness of fit of each spectra.

\subsection{Physical Property Measurements}

Magnetization measurements were performed in a Quantum Design Magnetic Property Measurement System (MPMS-3) SQUID magnetometer operating in the VSM mode. For the measurements at 300--600 K, the sample was mounted on a heater stick with alumina cement, and radiation shielding temperature homogeneity across the sample was ensured by a Cu foil. For the measurements conducted at 1.8--300 K, the sample was mounted on a quartz sample holder with GE 7031 varnish. The temperature dependent measurements were carried out using a 0.5 K/min warming/cooling rate.

Resistance measurements were performed using a Quantum Design Physical Property Measurement System (PPMS) in AC transport mode. The samples were prepared by cutting the crystals into rectangular bars, and the contacts were made by spot welding 25 $\mu$m thick annealed Pt wire onto the samples in standard four point geometry so that the current was applied within the \textit{ab}-plane. To ensure good mechanical strength, a small portion of silver epoxy was painted over the spot-welded contacts, and typical contact resistances were $\approx$ 1 $\Omega$.  The transverse magnetoresistance was measured up to 90 kOe with the field applied perpendicular to the direction of current flow.

\subsection{Electronic Band Structure Calculations}

Band structure, density of states, and total energy for CrPt$_5$P were calculated in density functional theory\autocite{hohenberg1964inhomogeneous,PhysRev.140.A1133} (DFT) using the PBEsol\autocite{PhysRevLett.100.136406} exchange-correlation functional with spin-orbit coupling (SOC) included. All DFT calculations were performed in VASP\autocite{PhysRevB.54.11169,KRESSE199615} with a plane-wave basis set and projector augmented wave method.\autocite{PhysRevB.50.17953} We used the unit cell of 7 atoms with a $\Gamma$-centered Monkhorst-Pack\autocite{PhysRevB.13.5188} (10 $\times$ 10 $\times$ 6) k-point mesh with a Gaussian smearing of 0.05 eV. The kinetic energy cutoff was 319 eV. The ferromagnetic moment direction was changed along different axis to find the preferred direction.
 
\section{Results and Discussion}
\subsection{Phase Determination and Crystal Structure of Cr$_{1+\text{x}}$Pt$_{5-\text{x}}$P}

% Table generated by Excel2LaTeX from sheet 'Sheet2'
\begin{table}[b]
  \small
  \centering
  \caption{Atomic Coordinates and Isotropic Atomic Displacement Parameters for Cr$_{1+x}$Pt$_{5-x}$P determined by Rietveld refinement of the PXRD pattern in Figure 
  \ref{Structure_PXRD}b. See the text for information on refinement of the Pt site occupancy. The refinement statistics are \textit{R}$_{\text{wp}}$ = 11.5 and GOF = 3.2.
}
    \resizebox{\linewidth}{!}{\begin{tabular}{llrrrrl}
    \toprule
    Atoms & Wycoff & \multicolumn{1}{l}{Occ.} & \multicolumn{1}{l}{\textit{x}} & \multicolumn{1}{l}{\textit{y}} & \multicolumn{1}{l}{\textit{z}} & \textit{U}$_{\text{iso}}$ (\AA$^2$) \\
    \midrule
    Pt1 & 1\textit{a}  & 1   & 0   & 0 & 0 & 0.006(1) \\
    Pt2 & 4\textit{i}  & 1   & 0   & \multicolumn{1}{l}{1/2}   & \multicolumn{1}{l}{0.2905(1)} & 0.006(2) \\
    Cr3   & 1\textit{c}  & 1   & \multicolumn{1}{l}{1/2} & \multicolumn{1}{l}{1/2} & 0 & 0.01 \\
    P4   & 1\textit{b}  & 1   & 0   & 0   & \multicolumn{1}{r}{1/2}   & 0.01 \\
    \bottomrule
    \end{tabular}}%
  \label{AtomsU}%
\end{table}%

\begin{figure*}[!t]
    \centering
    \includegraphics[width=\linewidth]{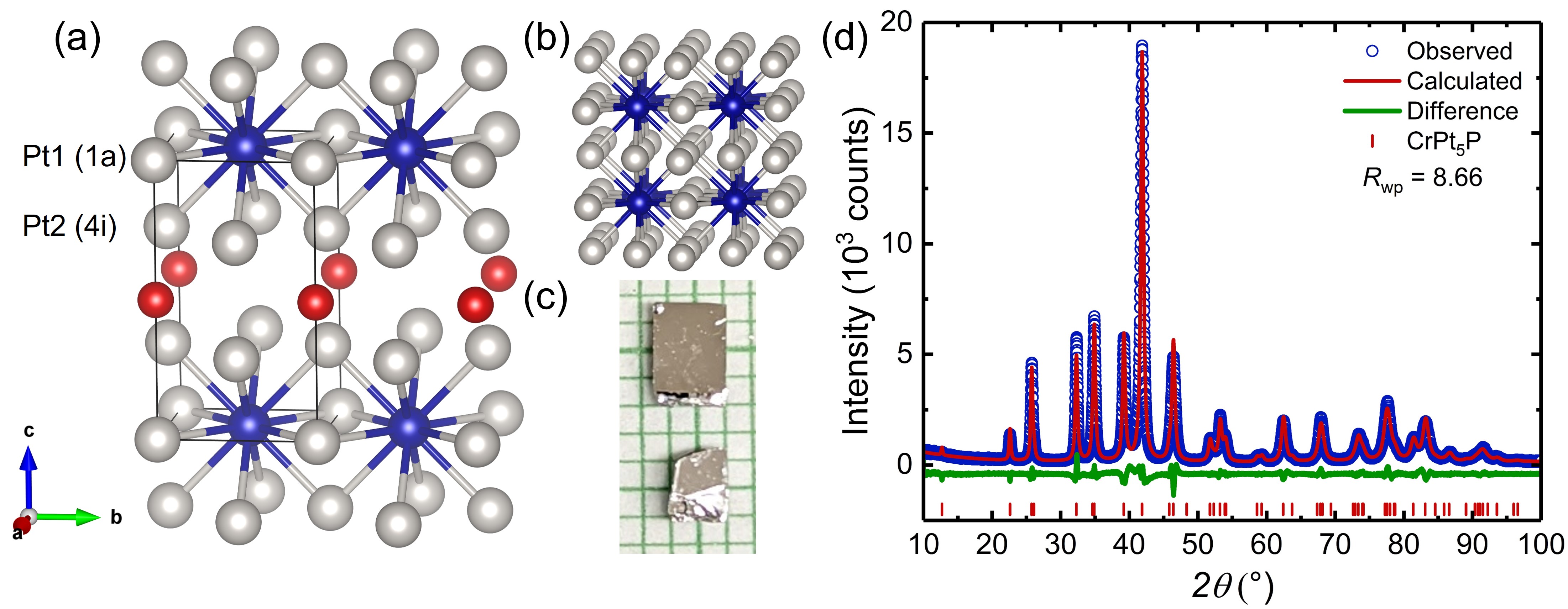}
    \caption[]{(a) Crystal structure of (a) CrPt$_5$P and (b) CrPt$_3$. The color code is blue = Cr, grey = Pt, and red = P. (c) Pictures of several CrPt$_5$P single crystals on a mm grid. (d) Powder X-ray diffraction pattern and Le Bail refinement for CrPt$_5$P.}
    \label{Structure_PXRD}
\end{figure*}

Cr$_{1+x}$Pt$_{5-x}$P (discussion of \textit{x} in following paragraphs) is a previously unknown 1-5-1 compound that adopts a layered tetragonal crystal structure with the space group \textit{P}4/\textit{mmm} ($\#$123), like the previously reported \textit{X}Pt$_5$P analogues.\autocite{el1970verzerrte,zakharova2018ternary,gui2020crystal,gui2020chemical,gui2021spin} The structure is illustrated in Figure \ref{Structure_PXRD}a and consists of slabs of face sharing CrPt$_{12}$ polyhedra that span the \textit{ab}-plane and which are separated along the \textit{c} axis by sheets of P atoms. As noted in the introduction, CrPt$_5$P can also be visualized as a composite built from single layers of CrPt$_3$ (crystal structure shown in Figure \ref{Structure_PXRD}b) in the Cu$_3$Au structure, which are separated along the \textit{c}-axis by Pt--P--Pt slabs. This second viewpoint provides a useful lens to view the magnetic properties.

Figure \ref{Structure_PXRD}d shows a powder X-ray diffraction pattern collected from several plate-like crystals. The Bragg peaks are somewhat broad compared to those collected on other 1-5-1 compounds,\autocite{Slade_Canfield_ZAAC} which is likely because the Cr$_{1+x}$Pt$_{5-x}$P plates are malleable and challenging to grind into a fine powder. Nevertheless, the experimental pattern agrees well with the reflections expected from the \textit{P}4/\textit{mmm} structure, and there are no significant peaks from secondary phases, indicating good sample purity. The lattice parameters determined from Le Bail refinements of the powder data are \textit{a} = 3.8922(2) \AA\ and \textit{c} = 6.8565(1) \AA\, which are very similar to those reported for MnPt$_5$P and FePt$_5$P.\autocite{gui2020chemical,gui2021spin} The Le Bail refinement statistics are \textit{R}$_{\text{wp}}$ = 8.66 and GOF = 2.39, indicating a reasonable refinement. 

\begin{figure*}[!t]
    \centering
    \includegraphics[width=0.8\linewidth]{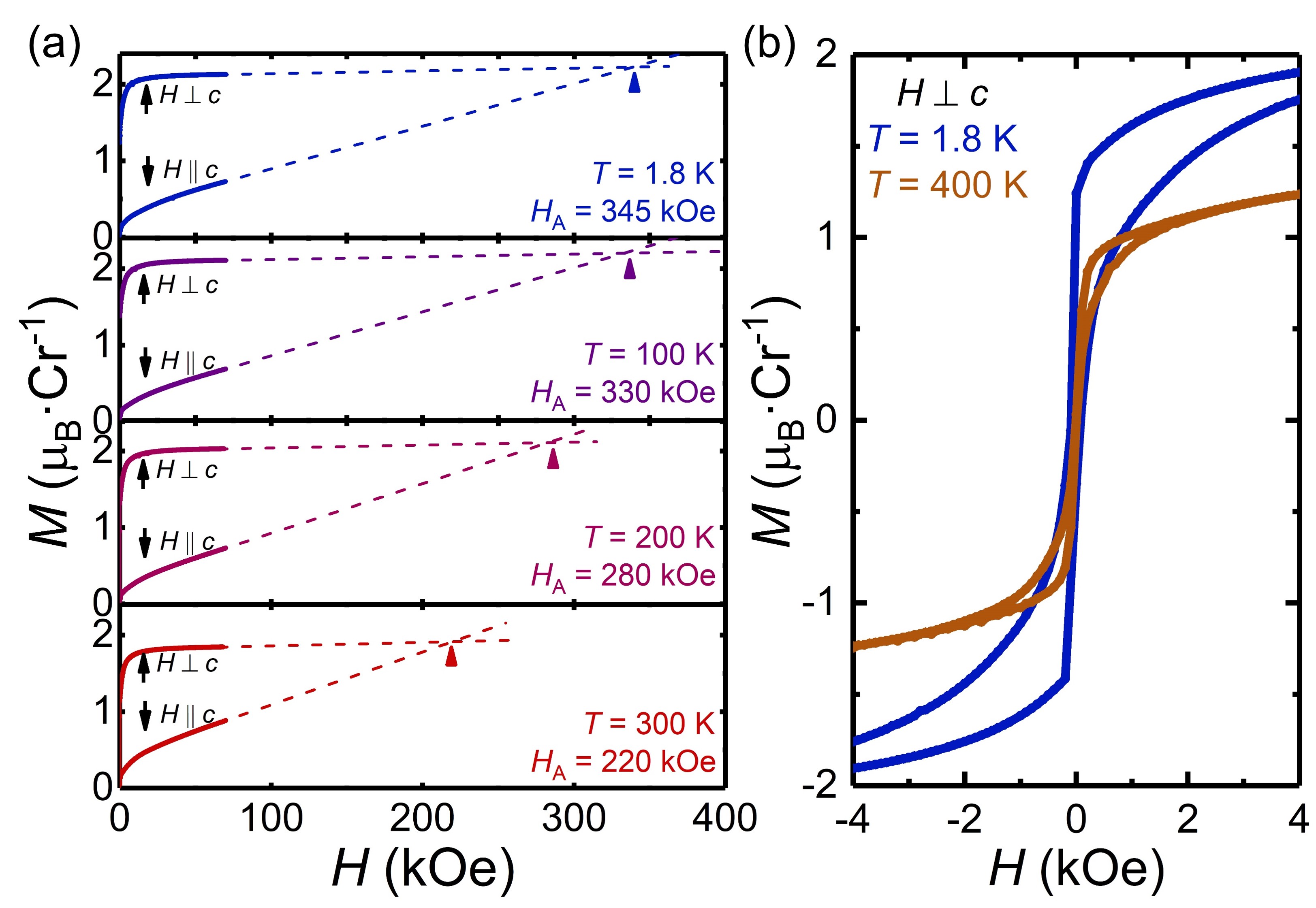}
    \caption[]{(a) Field dependent magnetization isotherms at temperatures between 1.8--300 K. The solid points are the data and the dashed lines are extrapolations to estimate the anisotropy fields. (b) Full magnetization isotherms in the easy direction (\textit{H} $\perp$ $\textit{c})$ at 1.8 K and 400 K showing hysteresis when raising/lowering the field.}
    \label{CrPt5P_MH}
\end{figure*}

\begin{figure}[!t]
    \centering
    \includegraphics[width=\linewidth]{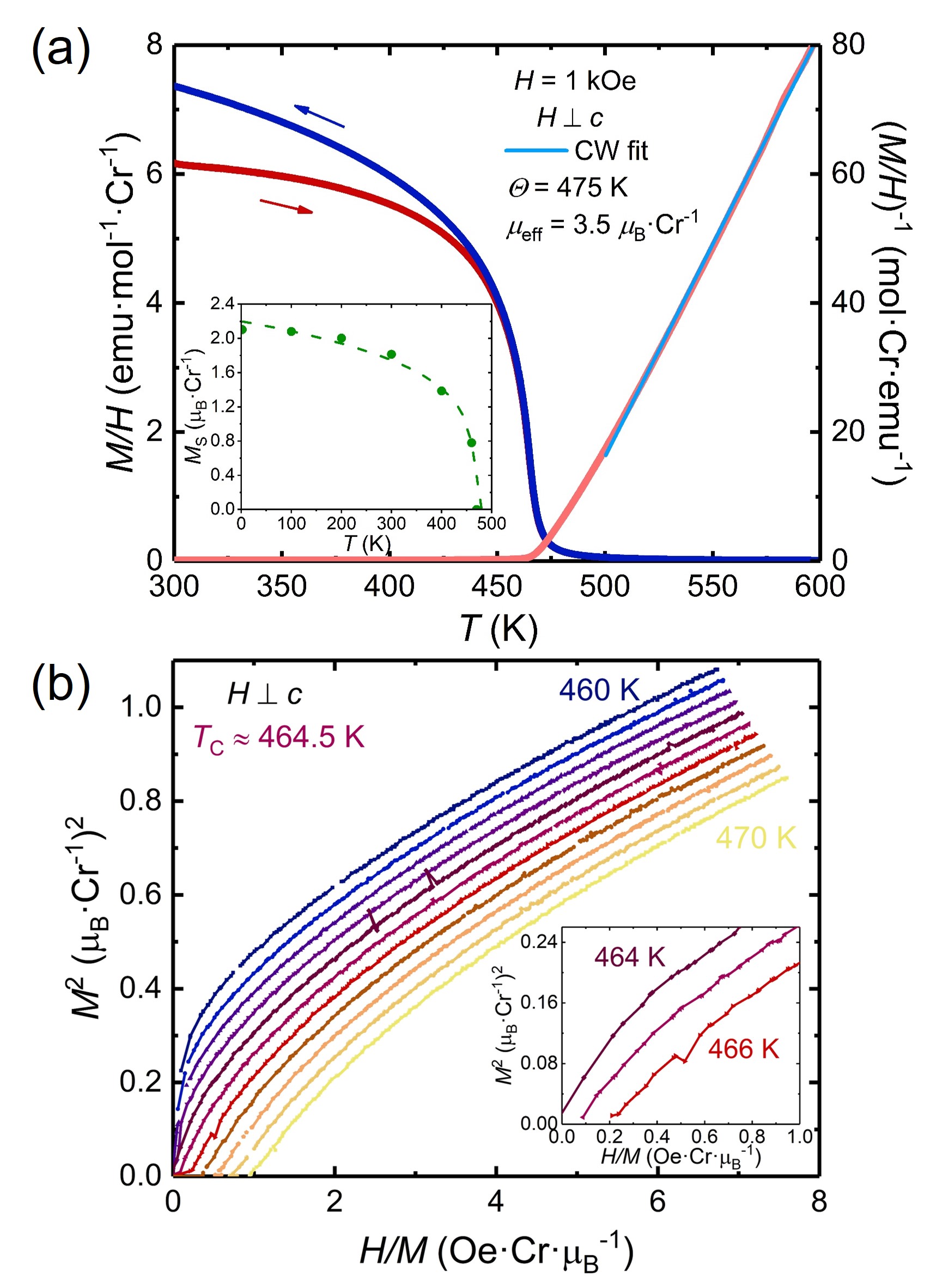}
    \caption[]{(a) Temperature dependence of \textit{M/H} (left axis) and \textit{M/H}$^{-1}$ (right axis). The light blue line marks the Curie-Weiss fit to the high temperature data. The inset shows the temperature dependence of the spontaneous moment \textit{M}$_{\text{s}}$ at select temperatures between 1.8-470 K, and the dashed line is an order-parameter fit (see text for more details). (b) Arrot plots between 460--470 K. The inset shows a closeup of the data between 464--466 K.}
    \label{CrPt5P_MT}
\end{figure}

Elemental analysis with EDS indicated that the samples are enriched with Cr and deficient in Pt compared to the expected stoicheometry of CrPt$_5$P, with true composition Cr$_{1.53(4)}$Pt$_{4.5(1)}$P$_{0.94(3)}$. The fact that the excess Cr is offset by a corresponding Pt deficiency, i.e. Cr$_{1+\text{x}}$Pt$_{5-\text{x}}$P, suggests there is mixed occupancy between Cr and Pt atoms where the Cr site is fully occupied and the extra Cr sits on the Pt sites. Considering that the parent compound CrPt$_3$ is known to form a solid solution with mixed occupancy between Cr and Pt atoms, where the Cu$_3$Au structure is maintained for compositions $\approx$20--40 $\%$ Cr,\autocite{besnus1973magnetic,goto1977effects} it is reasonable to believe that Cr$_{1+\text{x}}$Pt$_{5-\text{x}}$P can also accommodate excess Cr atoms on the Pt sites. Unfortunately, because the samples are malleable and easily deformed, we were unable to select a suitable, strain-free, piece for single crystal X-ray refinement (see experimental section), and the Bragg peaks were always smeared/broadened. Consequentially, we were not able to satisfactorily assess the site occupancy using single crystal refinements. We next attempted Rietveld refinements of our PXRD data using the known atomic positions of the structural analogue MnPt$_5$P as starting parameters.\autocite{gui2020chemical} Refining the Pt occupancy \textit{f} gives \textit{f}$_{Pt}$ $\approx$ 0.9 for both Pt sites, which is consistent with partial replacement of the Pt atoms with substantially lower \textit{Z} Cr, in qualitative agreement with the Cr$_{1.5}$Pt$_{4.5}$P composition implied by the EDS analysis. We note that the thermal displacement parameters \textit{U} for Cr and P were fixed to 0.01 in order to ensure a stable refinement. This is likely necessary given that the large and heavy Pt atoms contribute more than 70$\%$ atomic weight in Cr$_{1+x}$Pt$_{5-x}$P and can significantly absorb X-rays. The refinements are otherwise consistent with those in other \textit{X}Pt$_5$P systems.\autocite{gui2020chemical,gui2021spin} Given the limitations of our powder data, we were unable to obtain a stable refinement when explicitly attempting to model the disorder by introducing Cr onto the two Pt sites, and therefore cannot currently determine whether the excess Cr preferentially occupies one of the two Pt sites. Ultimately, our PXRD and EDS results both provide clear evidence for mixed occupancy between Cr and Pt atoms on the Pt-sublattice; however, quality single crystal diffraction data or high resolution synchrotron PXRD is needed for fully quantitative structural characterization.

\subsection{Magnetic Properties of Cr$_{1+\text{x}}$Pt$_{5-\text{x}}$P}

Figure \ref{CrPt5P_MH}a shows magnetization isotherms temperatures collected from 0--70 kOe and at temperatures between 1.8--300 K. Measurements were conducted on a plate-like crystal with EDS composition of Cr$_{1.5}$Pt$_{4.5}$P, and the field was applied both along the \textit{c}-axis and within the \textit{ab}-plane (\textit{H} $\perp$ \textit{c}). The \textit{M}(\textit{H}) results indicate Cr$_{1.5}$Pt$_{4.5}$P is ferromagnetic below at least 300 K with strong easy-plane anisotropy. As will be shown below, the Curie temperature is $\approx$ 465 K. Full \textit{M}(\textit{H}) loops obtained at 1.8 K and 400 K are displayed in Figure \ref{CrPt5P_MH}b and show clear hysteresis between increasing and decreasing field sweeps, confirming the ferromagnetic nature of Cr$_{1.5}$Pt$_{4.5}$P. The coercive field is $\approx$ 0.2 kOe, indicating the ferromagnetism is rather soft, similar to what is observed in other 1-5-1 ferromagnets MnPd$_5$P and MnPt$_5$As, which both have small coercive fields under 50 Oe.\autocite{Slade_Canfield_ZAAC,gui2020crystal,gui2020crystal} In the \textit{H} $\perp$ \textit{c} orientation (i.e. with the field applied in the basal plane), the magnetization increases in a nearly step-like fashion at low fields, quickly reaching saturation by $\approx$ 4 kOe, and the saturated moment falls monotonically with increasing temperature, from 2.1 $\mu_{\textit{B}}$/Cr at 1.8 K to 1.8 $\mu_{\textit{B}}$/Cr at 300 K. At 1.8 K, the saturated moment of $\approx$ 2.1 $\mu_{\text{B}}$ is very close to the expected 2 $\mu_{\text{B}}$ for \textit{S} = 3/2 Cr$^{3+}$. 

In the magnetic hard direction (\textit{H} $\parallel$ \textit{c}), the magnetization shows a small increase at the lowest fields and settles into a linear field dependence as \textit{H} approaches 70 kOe, reaching a maximum of $\approx$ 0.7 $\mu_{\text{B}}$ at 1.8 K, which is only a third of the saturated 2.1 $\mu_{\text{B}}$ measured when the field is applied in the \textit{ab}-plane. To determined the anisotropy fields \textit{H}$_A$, we extrapolated the tangents of the \textit{H} $\parallel$ \textit{c} and \textit{H} $\perp$ \textit{c} curves to locate the field in which the respective lines meet, as shown in Figure \ref{CrPt5P_MH}a, giving an estimated \textit{H}$_A$ of $\approx$ 345 kOe at 1.8 K that falls monotonically with increasing temperature to $\approx$ 220 kOe at 300 K. Note that all of our \textit{M}(\textit{H}) measurements in the hard direction (\textit{H} $\parallel$ \textit{c}) show a small low-field saturation below $\approx$ 2 kOe, followed by more linear behavior at high fields. This is most likely from to a small misorientation of the sample (or part of the sample, due to malleability) such that the field is not applied exactly along the \textit{c}-axis; however, at this time we cannot fully rule out the possibility of a small axial component to the ferromagnetism. Assuming the low-field saturation is from imperfect sample orientation, the anisotropy fields discussed above will slightly underestimate the true values.

Figure \ref{CrPt5P_MT}a shows the temperature dependence of the magnetization (\textit{M/H}) measured in a \textit{H} = 1 kOe applied field for Cr$_{1.5}$Pt$_{4.5}$P between 300--600 K. On cooling, \textit{M/H} increases rapidly beginning near 470 K, characteristic of the onset of ferromagnetic order. The right axis of Figure \ref{CrPt5P_MT}a shows the inverse (\textit{M/H})$^{-1}$ of the temperature dependent \textit{M/H} data above 300 K. (\textit{M/H})$^{-1}$ has a nearly linear, Curie-Weiss like temperature dependence above $\approx$ 500 K. Curie-Weiss fits to the \textit{H/M}$^{-1}$ data above 500 K give an estimated Weiss temperature $\Theta$ = 475 K, where the positive value is consistent with ferromagnetic ordering, and an effective moment $\mu_{\text{eff}}$ = 3.5 $\mu_{\text{B}}$/Cr, which like the 1.8 K saturated moment discussed above, is in reasonably agreement with the 3.87 $\mu_{\text{B}}$ anticipated for paramagnetic Cr$^{3+}$ moments.

Because the applied field will broaden and enhance a ferromagnetic transition, the Curie temperature is poorly defined in temperature dependent magnetization data. To more accurately determine \textit{T}$_C$, we also collected magnetization isotherms at temperatures above 400 K and assembled Arrot plots.\autocite{PhysRev.108.1394} Figure \ref{CrPt5P_MT}c and its inset show the data collected every 1 K between 460 K and 470 K. Here, the plot of \textit{M}$^2$ against \textit{H/M} should run through the origin at the Curie temperature, and our results indicate \textit{T}$_C$ $\approx$ 464.5 K.

The inset to Figure \ref{CrPt5P_MT}a shows the magnitude of the spontaneous magnetization (\textit{M}$_{\text{s}}$), estimated by extrapolating the tangents of the high field \textit{M}(\textit{H}) data to zero field, increases in an order-parameter like manner as the temperature is lowered, approaching 2.1 $\mu_{\text{B}}$/Cr at 1.8 K. Linear fits to a plot of log(\textit{M}$_{\text{s}}$) vs. log(\textit{T}$_C$--\textit{T}) give a slope $\beta$ $\approx$ 0.23. The dashed line in the inset to Figure \ref{CrPt5P_MT}a shows the $(T_C-T)^{\beta}$ captures the experimental data well, and the estimated $\beta$ $\approx$ 0.23 suggests the magnetism in CrPt$_5$P is not mean-field like (where $\beta$ = 0.5).

\subsection{Transport properties of Cr$_{1+\text{x}}$Pt$_{5-\text{x}}$P}

\begin{figure}[!t]
    \centering
    \includegraphics[width=\linewidth]{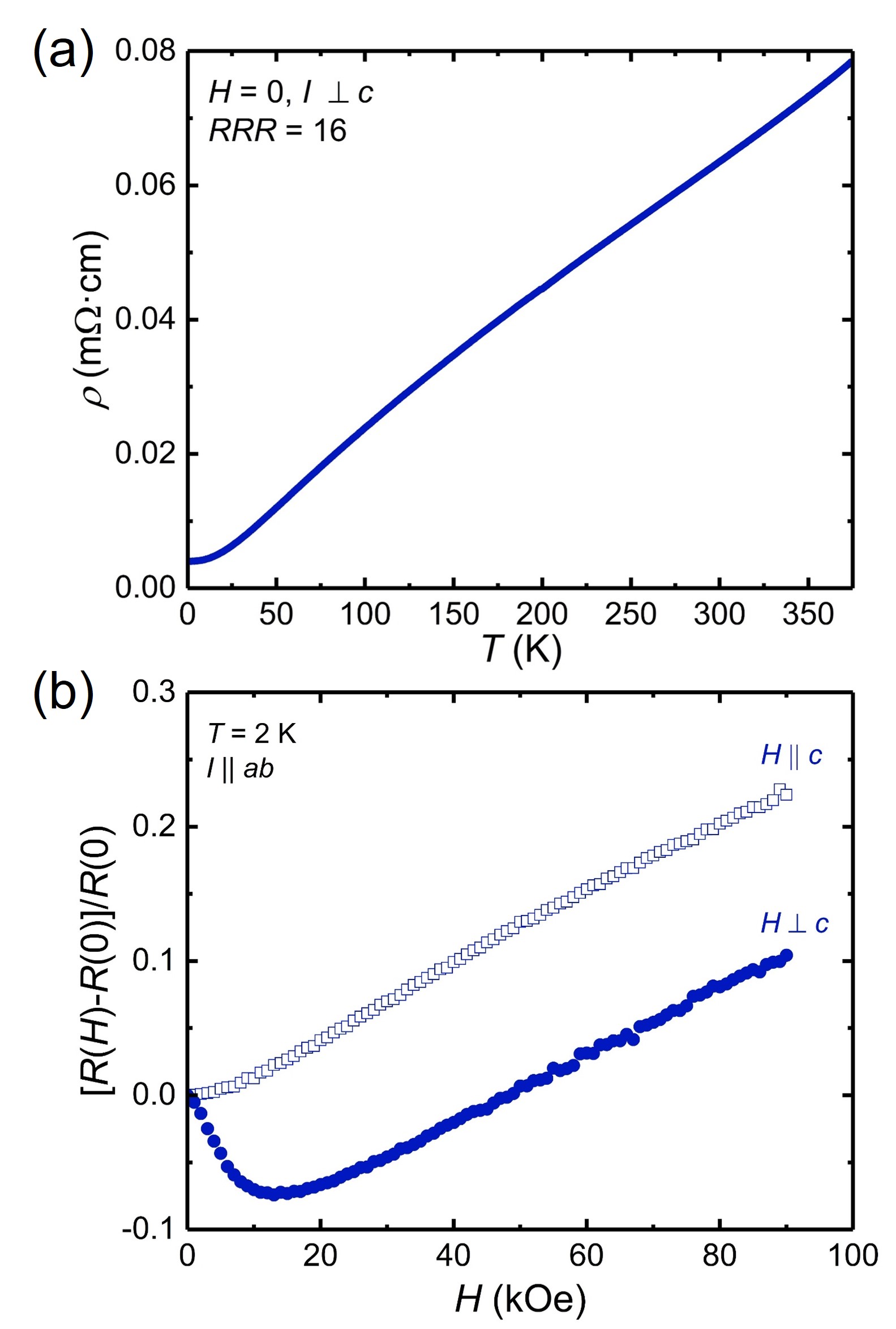}
    \caption[]{(a) Temperature dependent resistance of CrPt$_5$P between 1.8--375 K. (b) Transverse magnetoresistance at 2 K.}
    \label{RT_RH}
\end{figure}

Figure \ref{RT_RH}a shows the temperature dependent resistivity $\rho$ between 1.8--375 K. As expected, the behavior is characteristic of a metal, decreasing monotonically with cooling and approaching saturation as the lowest temperature is reached. No transitions are observed below 375 K, which is consistent with the single \textit{T}$_C$ = 464.5 K transition detected in the magnetic data. The residual resistance ratio \textit{RRR} = $\rho$(300 K)/$\rho$(1.8 K) is 16, indicating reasonably high crystal quality. Considering the mixed occupancy between Cr and Pt atoms inferred from the EDS and X-ray diffraction analysis, 16 may appear a high \textit{RRR} for a compound with such a substantial degree of crystallograrphic disorder ($\approx$ 10 $\%$ of the Pt sites are occupied by excess Cr); however, the \textit{RRR} of 16 is considerably lower than the \textit{RRR} $\approx$ 110 reported for single crystals of the structural analogue MnPt$_5$P.\autocite{Slade_Canfield_ZAAC} Using a similar effective temperature below the Neel temperature still gives \textit{RRR} $\approx$ 70 for MnPt$_5$P, suggesting the atomic disorder in Cr$_{1.5}$Pt$_{4.5}$P raises the residual resistance compared to what would be expected in an ideal, disorder-free crystal.

Figure \ref{RT_RH}b shows the transverse magnetoresistance, defined as [\textit{R}(\textit{H})--\textit{R}(0)]/\textit{R}(0)), collected at 2 K with the field applied both along the \textit{c}-axis and within the \textit{ab}-plane. In both cases, the current flow was within the plane and the field perpendicular to the direction of the current. At 2 K and for \textit{H} $\parallel$ \textit{c}, CrPt$_5$P has a relatively small positive magnetoresistance that reaches 20 $\%$ at 9 T. When the field is applied in the \textit{ab}-plane (the magnetic easy direction), the 2 K magnetoresistance first has a negative field dependence at low \textit{H} that crosses over to a positive slope near $\approx$ 15 kOe and then increases with a similar field dependence as the \textit{H} $\parallel$ \textit{c} data over the remaining range of \textit{H}. The negative slope at low fields is typical of ferromagnets and most likely comes from the decrease in spin disorder scattering as the field aligns the magnetic domains before crossing over to positive behavior at higher \textit{H}.

\subsection{Discussion}

Whereas the high Pt content and planar anisotropy likely limit the feasibility of CrPt$_5$P for practical applications, the 220 kOe (at 300 K) and 345 kOe (at 1.8 K) anisotropy fields are exceptionally high. For comparison, Nd$_2$Fe$_{14}$B, which is among the most widely used permanent magnets, has an anisotropy field of 82 kOe near room temperature.\autocite{bolzoni1987first} Other examples include MnBi (\textit{H}$_a$ = 50 kOe),\autocite{PhysRevB.90.174425} HfMnP (\textit{H}$_a$ = 100 kOe),\autocite{lamichhane2016discovery} FePt (\textit{H}$_a$ = 100 kOe),\autocite{inoue2006temperature} and CoPt (\textit{H}$_a$ = 140 kOe).\autocite{shima2005magnetocrystalline} Clearly the anisotropy of CrPt$_5$P is substantially stronger. As magnetic anisotropy arises from spin orbit coupling, it is very likely the substantial quantity of large-\textit{Z} Pt atoms in CrPt$_5$P underpins the high anisotropy. 

The 464.5 K Curie temperature of Cr$_{1.5}$Pt$_{4.5}$P is a high ordering temperature for a compound with only 1/7 moment bearing Cr atoms per formula unit. This likely indicates that the Cr atoms do not behave as local moments and that CrPt$_5$P is better described as an itinerant ferromagnet in which the magnetism arises from the exchange-splitting of the conduction electrons.\autocite{stoner1938collective,wohlfarth1951xliii,landrum2000orbital} A simple experimental metric for evaluating the degree of itinerancy in a metallic compound is the Rhodes-Wohlfarth ratio \textit{q}$_c$/\textit{q}$_s$,\autocite{rhodes1963effective} which is calculated from the high temperature effective moment and base temperature saturated moment using the following expressions:

\begin{center}
    
\(\mu_{eff}^2 = q_{c}(q_{c}+2)^2\mu_{B}\)

\(\mu_{sat} = q_{s}\mu_{B}\)

\(q_{c}/q_{s} = (-1+ \sqrt{1+(\mu_{eff}/\mu_{B})^2})/(\mu_{sat}/\mu_{B})\)

\end{center}

For compounds with relatively low magnetic ordering temperatures, high ratios suggest itinerant behavior, and \textit{q}$_c$/\textit{q}$_s$ should be close to unity when the magnetism is local moment like (i.e. in rare-earth containing materials). The Rhodes-Wohlfarth ratio also converges towards 1 in itinerant magnets when the ordering temperatures is high ($>$ 300 K).\autocite{takahashi2013spin} For Cr$_{1.5}$Pt$_{4.5}$P, we find \textit{q}$_c$/\textit{q}$_s$ = 1.3, which is consistent with the expectation for an itinerant magnetic with a relatively high, 465 K, Curie temperature. Because \textit{q}$_c$/\textit{q}$_s$ of 1 are also expected for local moment systems, which may be consistent with the trivalent oxidation state of Cr implied by the magnetic measurements, this result is ultimately somewhat ambiguous; however, the band structure calculations discussed below also support itinerant, Stoner-like magnetism.

\begin{figure}[!t]
    \centering
    \includegraphics[width=\linewidth]{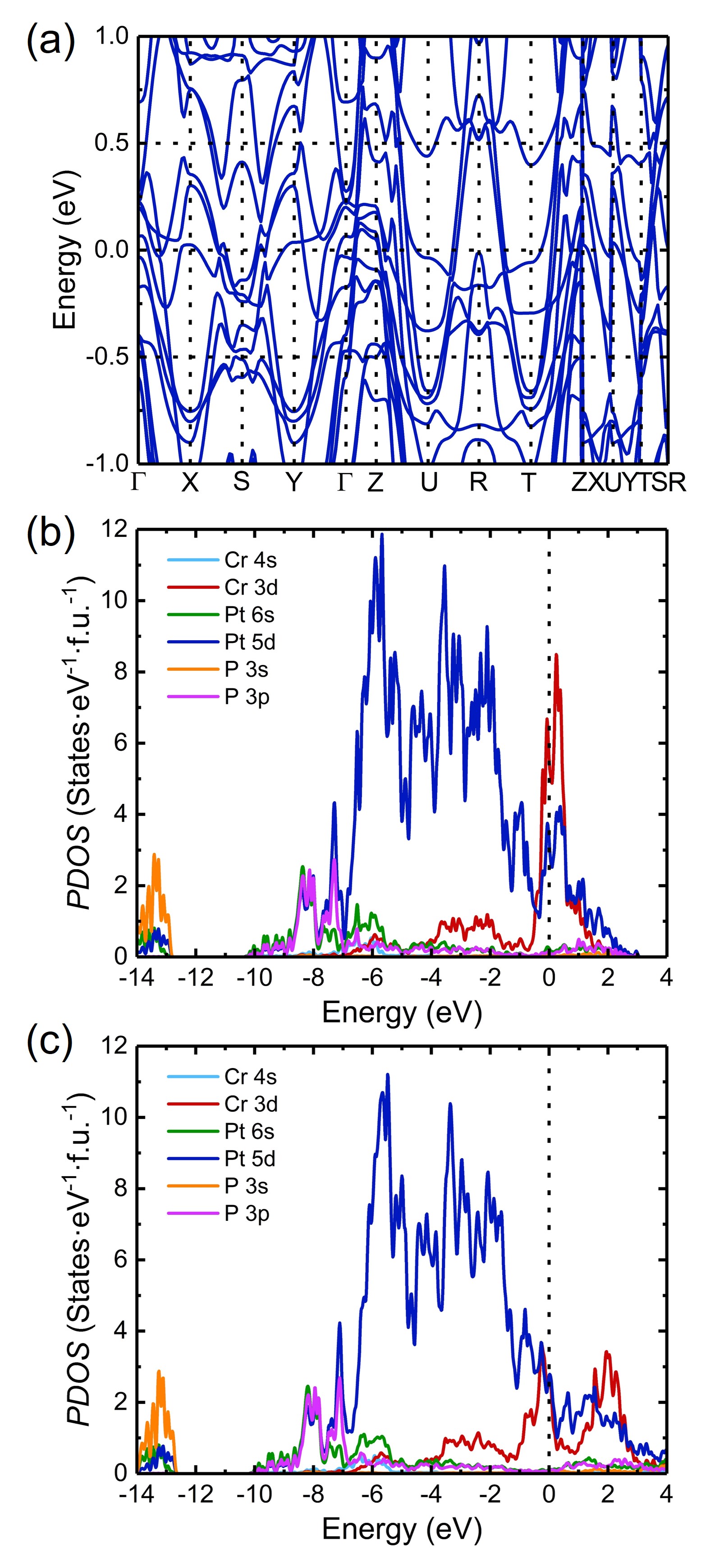}
    \caption[]{(a) Electronic band structure for CrPt$_5$P in the ferromagnetic phase (with moments along the \textit{a}-axis. (b) Density of states in the non-magnetic phase. (c) Density of states in the ferromagnetic phase. The calculations assume the ideal 1-5-1 composition (no Cr-Pt site disorder)}
    \label{ElectronicStructure}
\end{figure}

To provide greater insight into the magnetism and strong anisotropy measured in CrPt$_5$P, we used density functional theory (DFT) to calculate the electronic band structure, which is displayed in Figure \ref{ElectronicStructure}a. Our calculations considered the ideal 1-5-1 composition, i.e. without excess Cr and mixed occupancy between Cr and Pt atoms. Consistent with the experimental results, the calculations show CrPt$_5$P to be a metal with many well dispersed bands crossing the Fermi level (\textit{E}$_F$). The orbital projected density of states (PDOS) for nonmagnetic and ferromagnetic (with the moment along the \textit{a}-axis) phases of CrPt$_5$P are respectively given in Figures \ref{ElectronicStructure}b and \ref{ElectronicStructure}c. In both cases, the majority of the states near \textit{E}$_F$ are derived from Cr-3d and Pt-5d orbitals. The main contributions from P orbitals are significantly deeper in energy, with a band of P-3p based states from -10 to -5 eV and the P-3s states spanning -14 to -13 eV.

Comparing the density of states calculations for nonmagnetic and ferromagnetic CrPt$_5$P, we observe a narrow peak in the DOS around the Fermi level in the paramagnetic phase. The spike in DOS is composed primarily of Cr-3d orbital states with a smaller but still substantial contribution from Pt-5d orbitals, suggesting the hybridization between Cr-3d and Pt-5d orbitals in CrPt$_5$P is strong. As magnetic anisotropy arises from spin orbit coupling, the substantial hybridization between Cr-3d and the high \textit{Z} Pt-5d states at \textit{E}$_F$ is consistent with the strong magnetic anisotropy observed experimentally. Once in the ferromagnetic state, the peak at \textit{E}$_F$ is split, where the lower spin majority band is approximately centered at \textit{E}$_F$ and the higher spin minority band at 2 eV. Both upper and lower peaks retain significant contributions from both Cr-3d and Pt-5d orbital states. The calculations shown in Figures \ref{ElectronicStructure}b and \ref{ElectronicStructure}c match what is anticipated for an itinerant ferromagnet, where a locally high density of states near \textit{E}$_F$ is split into upper and lower, majority and minority, spin bands by the on-site Coulomb repulsion, giving rise to a net magnetization associated with the conduction electrons.\autocite{landrum2000orbital,stoner1938collective} Our calculations therefore support CrPt$_5$P as an itinerant ferromagnet, which is consistent with its Rhodes-Wohlfarth ratio and high Curie temperature.

Integrating the contributions of the spin up and down electrons, we calculated a magnetic moment of 3.2 $\mu_{\text{B}}$/Cr, which is in good agreement with the experimental $\mu_{\text{eff}}$ $\approx$ 3.5 $\mu_{\text{B}}$/Cr inferred from the Curie-Weiss fits. Moreover, the calculations indicate the most favorable ferromagnetic configuration is with the ordered moment within the \textit{ab}-plane, consistent with the experimental observations, and we calculate a magnetic anisotropy energy -1.261 meV/f.u. Although this suggests weaker anisotropy than the experimental results, the DFT calculations overall satisfactorily capture the magnetism experimentally observed in CrPt$_5$P.

Because the Cr$_{1+x}$Pt$_{5-x}$P studied here deviate from the ideal stoicheometry, with \textit{x} $\approx$ 0.5, we also constructed a (2x2x1) supercell with two Pt sites, one at 4\textit{i} and the other at 1\textit{a}, substituted by Cr, in order to preliminarily study the effect of excess Cr on the electronic structure. The calculations indicate that excess Cr placed on either Pt sites prefers to couple antiferromagneticallly with the host Cr; however, the additional Cr changes the DOS away from \textit{E}$_{\text{F}}$ at +2.0 eV and -1.0 eV, and we find that the DOS near \textit{E}$_{\text{F}}$ is not strongly altered by the presence of excess Cr. Consequentially, it is unclear from these calculations how strongly the excess Cr will alter the magnetism. In our experimental \textit{M}(\textit{H}) data, we observe no evidence for a ferrimagnetic state associated with AFM coupling between Cr moments associated with the different crystallographic positions, and the saturated moments are in reasonable agreement with the expectations for trivalent Cr, both suggesting Cr$_{1.5}$Pt$_{4.5}$P shows simple ferromagnetic order. A more detailed study on the effect of mixed occupancy requires configurational thermodynamics using cluster expansion method or coherent potential approximation, ideally in tandem with measurements on crystals with different compositions (\textit{x}).

As described in the discussion of the crystal structure, Cr$_{1+x}$Cr$_{5-x}$P can be envisioned as single layers of CrPt$_3$ that span the \textit{ab}-plane and which are divided along the \textit{c}-axis by planes of P atoms. From this perspective, it is interesting to compare the magnetic properties of Cr$_{1+x}$Cr$_{5-x}$P with those of the parent compound CrPt$_3$. In CrPt$_3$, the Cr atoms also order ferromagnetically with a Curie temperature of $\approx$ 490 K.\autocite{pickart1963neutron,besnus1973magnetic,maruyama1995separation,leonhardt1999crpt} In Cr$_{1+x}$Cr$_{5-x}$P, the insertion of the layer of P atoms nearly doubles the Cr-Cr distance along the \textit{c}-axis from 3.874 \AA\ in CrPt$_3$ to 6.857 \AA\ in CrPt$_5$P, whereas the in-plane Cr-Cr distances (the \textit{a}-lattice parameters) are approximately the same in each compound. Considering that the salient structural feature of face sharing Cr-Pt polyhedra is conserved in both compounds, and that the electronic states near the Fermi level in CrPt$_5$P are essentially derived only from Cr and Pt based orbital states, the similar magnetic properties suggests that the underlying physics governing the magnetism in each compound may be similar and that the addition of the P-layer in Cr$_{1+x}$Cr$_{5-x}$P serves to weaken the exchange interaction between magnetic moments and slightly lower the ordering temperature. As Cr$_{1+x}$Cr$_{5-x}$P and CrPt$_3$ share similar crystallographic and magnetic properties, investigation of topological transport properties, like the high anomalous hall effect found in CrPt$_3$, may be an attractive future direction for work on Cr$_{1+x}$Cr$_{5-x}$P.

\section{Summary and Conclusions}

We grew single crystals of Cr$_{1+x}$Pt$_{5-x}$P (\textit{x} = 0.5) and found that it is a remarkably anisotropic ferromagnet with a \textit{T}$_C$ = 464.5 K and a 1.8 K anisotropic field of 345 kOe. We show that single crystals can readily be grown out ternary Cr-Pt-P solutions. Like all other members of the \textit{X}(Pt,Pd)$_5$P family, Cr$_{1+x}$Pt$_{5-x}$P crystallizes in the \textit{P}4/\textit{mmm} space group and is comprised of CrPt$_3$-like layers spaced out by sheets of P atoms along the \textit{c}-axis. Our EDS and XRD analysis indicate Cr-enrichment, likely pointing toward site occupancy disorder and true compositions of Cr$_{1.5}$Pt$_{4.5}$P, which is similar to the disorder found in the parent compound CrPt$_3$. Likely owing to the high Pt content, CrPt$_5$P has very strong magnetic anisotropy in which the \textit{ab}-plane is the easy direction. The estimated anisotropy fields are $\approx$ 345 kOe and 220 kOe respectively at 1.8 K and 300 K. Cr$_{1.5}$Pt$_{4.5}$P exhibits metallic transport behavior and has relatively weak transverse magnetoresistance. Electron band structure calculations are consistent with itinerant magnetism, showing the CrPt$_5$P has a local peak in the density of states near the Fermi level that is split into upper and lower bands in the ferromagnetic state. Likewise, our calculations suggest strong hybridization between Cr-3d and Pt-5d states, in agreement with the experimentally observed strong magnetic anisotropy.

\vskip 0.25cm
\noindent
\textbf{\textit{Acknowledgements}}

Work at Ames National Laboratory was supported by the U.S. Department of Energy, Office of Science, Basic Energy Sciences, Materials Sciences and Engineering Division. Ames National Laboratory is operated for the U.S. Department of Energy by Iowa State
University under Contract No. DE-AC02-07CH11358. TJS, PCC, and LLW were supported by the Center for Advancement of Topological Semimetals (CATS), an Energy Frontier Research Center funded by the U.S. Department of Energy Office of Science, Office of Basic Energy Sciences, through Ames National Laboratory under its Contract No. DE-AC02-07CH11358 with Iowa State University. W.X. is supported by NSF-DMR-2053287.

\vskip 0.25cm
\noindent
\textbf{\textit{Conflicts of Interest}}

The authors have no conflicts of interest to declare.

\vskip 0.25cm
\noindent
*corresponding authors' email: slade@ameslab.gov, canfield@ameslab.gov

\printbibliography

\end{singlespace}

\end{document}